\documentclass[12pt, a4paper]{article}
\usepackage{graphicx}
\usepackage{multirow}
\usepackage{amsmath,amssymb,amsfonts}
\usepackage{amsthm}
\usepackage{mathrsfs}
\usepackage[title]{appendix}
\usepackage{xcolor}
\usepackage{textcomp}
\usepackage{authblk}
\usepackage{booktabs}
\usepackage{algorithm}
\usepackage{algorithmicx}
\usepackage{algpseudocode}
\usepackage{listings}
\usepackage{subcaption}
\usepackage{microtype}
\usepackage{hyperref}
\usepackage[margin=1in]{geometry}

\theoremstyle{plain}

\theoremstyle{definition}

\title{Approximate Hamiltonian Simulation Algorithm for Efficient Fluid Quantum Simulations}

\author{
    Zhiyuan Zhang,$^{1,2}$ Bolin Zhang,$^{1,2}$ Yongguang Lv,$^{1,2}$ Ruiqing He,$^{3}$ Hengliang Guo,$^{1}$ Jiandong Shang,$^{1,*}$ and Qiang Chen$^{1,\dag}$\\
    \small $^{1}$National Supercomputing Center in Zhengzhou, Zhengzhou University, Zhengzhou, Henan 450001, China \\
    \small $^{2}$School of Computers and Artificial Intelligence, Zhengzhou University, Zhengzhou, Henan 450001, China \\
    \small $^{3}$School of Communication and Artificial Intelligence, School of Integrated Circuits, Nanjing Institute of Technology, Nanjing, Jiangsu 211167, China
}
\date{}

\footnotetext{Corresponding author: sjd@zzu.edu.cn}
\footnotetext{Principal corresponding author: qiangchen@zzu.edu.cn}

\begin{document}

\maketitle
\begin{abstract}
This work aims to address the bottleneck issues of hardware resource limitation and decoherence error in the Hamiltonian simulation of quantum fluids, which are caused by the standard quantum Fourier transform and the evolution of momentum operators, resulting in excessively deep circuits and excessive two-qubit gates. We propose an approximate operator optimization scheme aimed at reducing the circuit depth in Hamiltonian evolution. The proposed scheme successfully reduces the depth of analog circuits from $O(n^2)$ to $O(nlogn)$ or even $O(n)$ by eliminating $O(n^2)$ redundant two-qubit entangling gates. In this work, the numerical experiments are implemented on a supercomputing-oriented quantum simulator, simulating two-dimensional unsteady divergent flow. Experimental results demonstrate that although the truncation of high-frequency qubit coupling terms introduces deterministic theoretical errors, scaling at $O(n)$ for AQFT and $O(n^2)$ for momentum truncation, the optimized simulations successfully preserve the inherent macroscopic temporal evolution characteristics of the fluid in a 10-qubit simulation, achieving high correlation coefficients of $r=0.933$, $r=0.941$, and $r=0.977$ for density, x-momentum, and y-momentum distributions respectively. Furthermore, we also analyzed the relationship between the algorithm truncation error and the hardware cumulative noise when the qubit number is extended to a higher level. This study proves that rationally adjusting truncation thresholds can establish an equilibrium point, preventing the hardware cumulative error from rapidly approaching 100\% at the 20-30 qubit scale, providing a feasible engineering pathway for simulating complex fluid systems on real quantum devices in the future.

\vspace{1em}
\noindent\textbf{Keywords:} Fluid Dynamics, Quantum Fluid Simulation, Approximation Operations, Hamiltonian Simulation, Approximate Quantum Fourier Transform
\end{abstract}

\newpage

\section{Introduction}\label{sec1}

Computational fluid dynamics using supercomputers has developed into a mature research tool, which has practical applications, such as in weather forecasting, shock wave analysis of hypersonic aircraft and simulating the flow of blood within the cardiovascular system. In order to efficiently solve nonlinear partial differential equations governing fluid motion on supercomputing platforms, researchers have developed and optimized a variety of classical numerical methods, the most representative of which include the finite volume method, which is widely used in engineering to deal with complex geometric boundaries, the finite difference method and the spectral method, which are suitable for high-precision calculation of regular grids. As well as the Reynolds-Averaged Navier-Stokes model, large eddy simulation and direct numerical simulation, which represent extremely high fidelity but consume a lot of computing power, derived from the challenge of high Reynolds number turbulence\cite{bib1,bib2}.

In quantum computing, the data of $n$ discrete grid points can be accurately mapped to the probability amplitude in the quantum superposition state by amplitude coding. Since n qubits can construct a Hilbert space containing $2^n$ basis States, this means that only $log_2n$ qubits are needed to fully represent the data of n classical grids. Thanks to the unique properties of quantum superposition and quantum entanglement, the use of quantum computing technology to study the problems to be solved by ordinary computing technology, such as how to solve the exponential computational cost, makes quantum computing provide a promising direction for the future development of fluid dynamics simulation.\cite{bib3,bib4,bib5}.

Fluid dynamics quantum simulation is an emerging frontier at the intersection of quantum computing and computational fluid dynamics. Its core objective is to leverage the advantages of quantum computers, such as parallelism, to address complex fluid problems—including high-dimensional and high-Reynolds-number turbulence—that are challenging for classical computers. Currently, there are two primary approaches to fluid dynamics simulation using quantum computing: hybrid quantum-classical algorithms \cite{bib6,bib7,bib8,bib9,bib10,bib11,bib12} and Hamiltonian simulation \cite{bib13,bib14,bib15,bib16,bib17,bib18,bib19}. The first method capitalizes on the highly parallelizable nature of quantum computation, but its efficiency is often hampered by frequent data exchange, state preparation, and measurement between the two types of computer hardware. The second approach, however, shows greater promise. It directly encodes the characteristics of fluid flow into a quantum system and then performs Hamiltonian evolution on a quantum processor, thereby avoiding the need for repeated quantum state preparation and measurement.

Michael Griebel fisrtly summarized the comprehensive theoretical framework of computational fluid dynamics from mathematical modeling to engineering implementation\cite{bib20},This work has laid the foundation for the numerical simulation of fluid mechanics. However, when facing complex multi-scale physical processes such as high Reynolds number turbulence, the grid resolution and computational cost of classical numerical simulation will increase dramatically with the increase of Reynolds number. This exponentially growing demand for computing power highlights the physical necessity and engineering urgency of exploring quantum computing. R. Steijl conducted a parallel evaluation study on quantum algorithms for computational fluid dynamics and developed a CFD solving framework based on a classical-quantum hybrid architecture, providing early parallel implementation schemes and performance evaluation benchmarks for quantum computing in numerical simulation of fluid dynamics\cite{bib21}. In practice, the frequent data exchange, quantum state preparation and measurement process between classical and quantum hardware will inevitably seriously drag down its operational efficiency. F. Gaitan proposed a general quantum algorithm for solving the Navier-Stokes equations\cite{bib22}. Based on the generalized Koopman–von Neumann formulation of classical mechanics, I. Joseph introduced a quantum simulation method for nonlinear classical dynamics, establishing a universal theoretical framework for quantum simulation adaptable to arbitrary nonlinear, non-Hamiltonian classical dynamical systems\cite{bib23}. This is a highly groundbreaking theoretical work that breaks the limitations of conventional Hamiltonian systems for quantum simulation and presents an extremely important mathematical approach that demonstrates that a wider range of general complex fluid mechanisms with nonlinear and non-Hermitian features can be treated with limited coherence time. Z.-Y. Chen addressed the computational efficiency bottleneck of classical finite volume methods in steady-state computational fluid dynamics by proposing a complete quantum-accelerated solving scheme, reconstructing the core solving process of the finite volume method through quantization, and providing a practical implementation path for the quantum upgrade of traditional CFD finite volume methods\cite{bib24}. W. Itani and S. Succi systematically analyzed the Carleman linearization for the incompressible fluid lattice Boltzmann equation with the BGK equilibrium function, laying the core theoretical foundation for constructing quantum lattice Boltzmann algorithms\cite{bib25}. P. Pfeffer proposed two hybrid quantum-classical reservoir computing reduced-order models\cite{bib26}. Z. Meng introduced a quantum computing framework for fluid dynamics based on the hydrodynamic Schrödinger equation and developed a corresponding predictor-corrector quantum solving algorithm\cite{bib27}. S. Succi and W. Itani addressed the high computational cost of simulating classical fluid systems by proposing a quantum computing platform-adapted method for fluid system simulation\cite{bib28}. That same year, they also introduced the QALB quantum algorithm for simulating incompressible fluid lattice Boltzmann with nonlinear BGK collision terms. By leveraging Carleman linearization and a bosonic mode coupling framework, they reconstructed both the collision and streaming core steps of LBM into unitary operator evolution forms, overcoming the limitations of classical Carleman techniques in neighboring variable coupling and achieving a fully quantum solution for nonlinear collision terms\cite{bib29}, By mapping the collision dynamics of microscopic particles into unitary operations through ingenious mathematical transformations, they successfully avoided the measurement overhead in the hybrid algorithm, which proved the feasibility of fully closed-loop calculation of nonlinear hydrodynamics in the quantum regime. In 2024, Zhaoyuan Meng \cite{bib30, bib31} established a mathematical mapping between the fluid Navier-Stokes equations and the quantum Schrödinger-Pauli equations via the generalized Madelung transform. On the algorithmic front, a Hamiltonian simulation algorithm based on fluid quantum representation was proposed, enabling quantum simulation of compressible/incompressible flows with vortex dynamics and scalar advection-reaction-diffusion problems \cite{bib32, bib33}.

However, in the NISQ era, Hamiltonian simulation faces severe challenges. For fluid dynamics algorithms based on the Schrödinger equation, the core step relies on the quantum Fourier transform, as it enables the conversion of quantum states between the computational basis and the phase basis—an indispensable procedure\cite{bib34}. However, the standard QFT involves \textit{O(n²)} two-qubit controlled-phase gates, and the number of such gates grows quadratically with the number of qubits\cite{bib35, bib36}. In the momentum operator evolution part, the squared wavenumber term involves global cross-interactions between all pairs of qubits. Both the QFT and the momentum operator evolution in the Hamiltonian require a large number of two-qubit gates. On actual superconducting quantum chips, only physically adjacent qubits can directly interact using two-qubit CZ gates\cite{bib37}. To forcibly implement interactions between non-adjacent qubits on hardware, a large number of SWAP gates must be inserted, making the overall circuit depth extremely deep\cite{bib38, bib39}. This simulation process accumulates gate errors and induces decoherence effects on real quantum devices, thereby affecting both the results and duration of the simulation\cite{bib40}.

To overcome the hardware limitations of quantum computing devices, this paper proposes an optimization framework based on approximate operations, which introduces controllable algorithmic errors in exchange for reduced circuit depth. We have made two key improvements to the original Hamiltonian simulation approach: first, replacing the standard Quantum Fourier Transform module with an Approximate Quantum Fourier Transform, and compensating for the approximation using single-qubit gates after the approximate Fourier transformation. By setting a threshold for rotation angles, we remove long-range controlled-phase gates that have negligible impact on the phase rotation, with the discarded phase shifts restricted to less than $2\pi/2^b$ based on the truncation threshold $b$, thereby reducing the number of two-qubit gates and mitigating gate-related errors. Second, during the evolution in momentum space, we introduce a momentum operator truncation strategy. For the momentum evolution operator $e^{-i\hat{k}_\alpha^2t/2}$, a truncation threshold is applied to eliminate two-qubit gates that meet the conditions, further compressing the depth of the simulation circuit and reducing resource consumption.

We implemented this approximate Hamiltonian simulation algorithm on the quantum simulator of the Songshan supercomputer at the National Supercomputing Center in Zhengzhou, using 10 qubits to perform fluid quantum simulations of two-dimensional non-stationary divergent flows. The simulated quantum processors features single-qubit gates with up to 99.97\% fidelity and two-qubit gates with 99.67\% fidelity. Experimental results demonstrate that although the substitution of AQFT and truncation of momentum operators introduce theoretical approximation errors, this scheme reduces the depth of simulation circuits, thereby reducing the uncompiled circuit depth and the number of two-qubit gates from $O(n^2)$ to $O(nlogn)$ or even $O(n)$. The simulation demonstrated that the captured features successfully preserved the original macroscopic evolution of the flow field, that the macroscopic physical properties of mass and momentum diffusion were preserved with minimal correlation loss, and qualitatively reproduced spatial flow fields with correlation coefficients $r \ge 0.933$ compared to ideal simulations. In addition, we analyze the variation of the algorithm truncation error and the hardware cumulative noise when the number of bits is extended to higher levels, and the dynamic relationship between the algorithm truncation error and the hardware cumulative noise. This work demonstrates the capability of approximate computation strategies to simulate complex fluids on resource-constrained quantum hardware and highlights the important role of such strategies in addressing quantum fluid simulation challenges.

This paper is organized as follows. Section 2 provides an overview of the mathematical and physical framework, including spatial discretization, quantum state encoding, and the construction of the fluid momentum operator. Section 3 elaborates on the core optimization strategy, specifically introducing the approximate quantum Fourier transform with phase compensation mechanism and the momentum operator truncation method. Section 4 shows the experimental results obtained on a quantum simulator, verifying the effectiveness of the proposed truncation strategy in simulating two-dimensional unsteady divergent flows and discussing the extension of the truncation error in terms of scalability. Section 5 explores the dynamic trade-off between the algorithmic truncation error and the consumption of physical hardware resources.

\section{Hamiltonian Simulation of Schrödinger Evolution}\label{sec2}

\subsection{Fluid Simulation Mapping}\label{sec:fluid}

Based on the generalized Madelung transformation, the fluid density is defined as
\begin{equation}
\rho \equiv \sum_{j=1}^{n_\psi} |\psi_j|^2.
\end{equation}
The fluid momentum is defined as
\begin{equation}
J \equiv \frac{i\hbar}{2m} \sum_{j=1}^{n_\psi} (\psi_j \nabla \psi_j^* - \psi_j^* \nabla \psi_j).
\end{equation}
In these simulations, without loss of generality, we set the reduced Planck constant and particle mass to unity, i.e., $\hbar=1$ and $m=1$. We conducted quantum simulations for the expansion of a two-dimensional unsteady divergent flow, which serves as a simplified model of nozzle flow in compressible potential flow. Initially, the flow field is uniformly distributed along the x-direction, with mass concentrated near $y=0$, and its initial density is
\begin{equation}
\rho(x, y, 0) = e^{-y^2}.
\end{equation}
The vorticity of this flow field is zero, allowing it to be encoded as a single-component wave function
\begin{equation}
\psi(x, y, 0) = e^{-y^2+ix}.
\end{equation}

\subsection{Spatial Discretization and Quantum State Encoding}\label{sec:spatial}
Consider a two-dimensional fluid defined on a periodic domain $x \in [-\pi, \pi]^2$, discretized into $2^{n_x} \times 2^{n_y}$ uniform grid points. The spatial coordinates are given by
\begin{equation}
x_k = -\pi + k\Delta x \quad \text{,} \quad y_l = -\pi + l\Delta y
\end{equation}
with
\begin{equation}
\Delta x = 2\pi/2^{n_x} \quad \text{,} \quad \Delta y = 2\pi/2^{n_y}
\end{equation}
representing the grid spacings. The fluid state is encoded in the component wave functions $\psi \equiv [\psi_1, \dots, \psi_{n_\psi}]^T$. Using $n = n_x + n_y$ qubits, the wave function components are expressed in the computational basis as:
\begin{equation}
|\psi_j(t)\rangle = \frac{1}{||\psi_j||_2} \sum_{l=0}^{2^{n_y}-1} \sum_{k=0}^{2^{n_x}-1} \psi_j(x_k, y_l, t)|k + 2^{n_x}l\rangle,
\end{equation}
Where $|k + 2^{n_x}l\rangle$ corresponds to the computational basis state at grid point $(x_k, y_l)$

\subsection{Schrödinger Evolution and Momentum Operator}\label{sec:schrodinger}

Fluid dynamics follows the Schrödinger equation under the Hamiltonian $H = -\nabla^2/2 + V$. When simulating fluids without non-conservative volume forces, the Hamiltonian can be simplified to $H = -(\partial_x^2 + \partial_y^2)/2$. The corresponding wave function evolution can be achieved without Trotter decomposition using the following formula:
\begin{equation}
e^{-iHt} = e^{i\partial_x^2t/2}e^{i\partial_y^2t/2} = U_x(t)U_y(t).
\end{equation}
Under the computational basis, the digitized unitary evolution operator along a specific direction is given by
\begin{equation}
U_\alpha(t) = \widehat{\text{QFT}}_\alpha^\dagger e^{-i\hat{k}_\alpha^2t/2} \widehat{\text{QFT}}_\alpha.
\end{equation}
The mathematical expression for the wave number operator $\hat{k}_\alpha$ mapped onto the qubit is:
\begin{equation}
\hat{k}_\alpha = -\frac{1}{2}\left(I_{2n_\alpha} + \sum_{j=1}^{n_\alpha} 2^{n_\alpha-j} \hat{Z}_j\right) + 2^{n_\alpha} \tilde{Z}_1.
\end{equation}
In this wave number operator formula, $\hat{Z}_j$ represents the Pauli Z operator acting on the $j$-th qubit, and $I_{2n_\alpha}$ denotes the $2n_\alpha \times 2n_\alpha$-dimensional identity matrix.

\section{The Truncation Operation in Quantum Circuits}\label{sec:method}

\begin{figure}[h]
\centering
\includegraphics[width=0.8\textwidth]{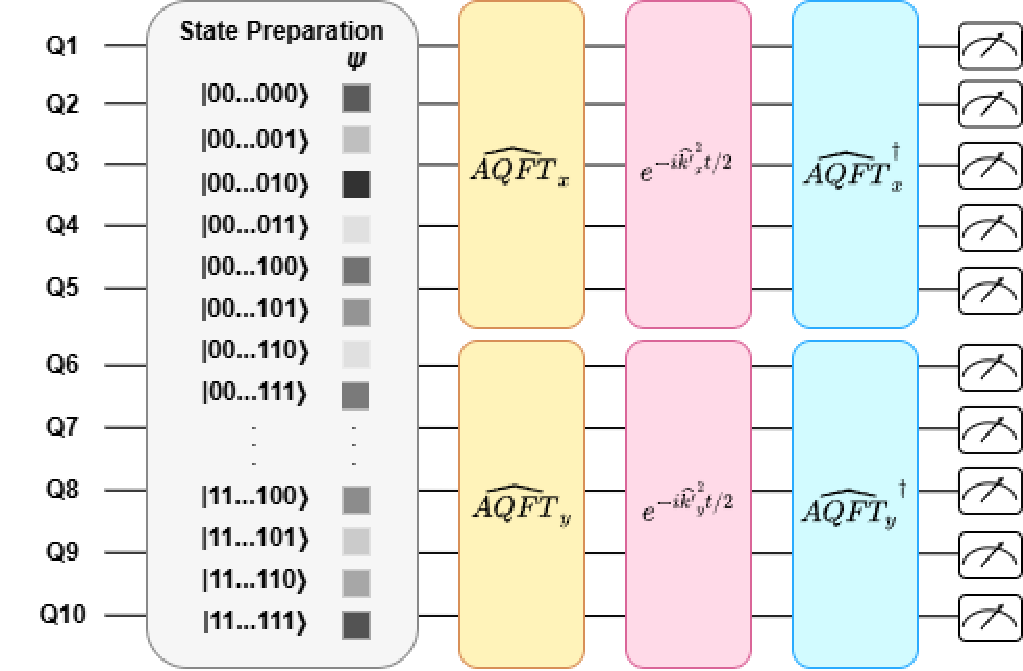}
\caption{The circuit schematic for the two-dimensional unsteady flow quantum simulation.}\label{fig:circuit}
\end{figure}

The circuit schematic for the two-dimensional unsteady flow quantum simulation in this study is illustrated in Figure~\ref{fig:circuit}. After encoding the wave function information into quantum bits, quantum state preparation is performed. The flow field state is then transformed into momentum space via an approximate quantum Fourier transform. A truncated and optimized unitary evolution operation is applied, followed by an inverse approximate quantum Fourier transform to revert the quantum state back to coordinate space, thereby completing the Hamiltonian simulation.

\subsection{Approximate Quantum Fourier Transform Based Quantum Circuit}

In standard Hamiltonian fluid simulations, the state preparation of evolution operators and momentum space transformations heavily rely on the quantum Fourier transform (QFT). For an $n$-qubit system, the standard QFT circuit requires $O(n^2)$ two-qubit controlled-phase gates ($CR_k$). To reduce the number of two-qubit gates in the circuit, we employ an approximate quantum Fourier transform (AQFT) to replace the standard QFT module. In this section, we incorporate truncation operations for two-qubit gates and compensation operations for the truncated gates.

\textbf{Truncation operation:} A standard QFT circuit consists of a series of controlled phase rotation gates $CR_k$, applied between the control qubit $q_c$ and the target qubit $q_t$, defined as:
\begin{equation}
CR_k=
\begin{pmatrix}
1 & 0 & 0 & 0 \\
0 & 1 & 0 & 0 \\
0 & 0 & 1 & 0 \\
0 & 0 & 0 & e^{2\pi i/2^k}
\end{pmatrix}
\end{equation}
where $k=|c-t|$ represents the index distance between the two qubits in the circuit.

As the index difference $k$ between the control and target qubits increases, the physical distance also grows, while the rotation angle $\theta = 2\pi/2^k$ of the $CR_k$ gate decreases exponentially. For large $k$, the minimal phase shift of $2\pi/2^k$ introduced by this $CR_k$ gate becomes completely overwhelmed by the inherent noise and decoherence effects of two-qubit gates in actual hardware. Moreover, due to the nearest-neighbor coupling limitation in real superconducting quantum processors, when implementing a long-range controlled-phase gate $CR_k$, the quantum compiler must automatically insert a series of SWAP gates to transport distant qubit states adjacent to the current qubit. After performing the controlled operation, these states must then be returned to their original positions. Consequently, retaining such gates not only fails to enhance algorithmic accuracy but also increases the overall circuit depth, leading to unnecessary resource consumption and reduced simulation fidelity.

We introduce a parameter $b$ as the truncation threshold. During the circuit compilation phase, we traverse all $CR_k$ gates in the standard QFT circuit. When $k \le b$, the controlled-phase gate is retained; when $k > b$, the controlled-phase gate is removed from the circuit. Prior to the aforementioned truncation operation, the circuit depth of the standard QFT was $O(n^2)$. After truncation, all $CR_k$ gates with physical distance $k > b$ and their associated SWAP gate networks are entirely removed.

For any $i$-th qubit, it is at most connected to qubits within a distance not exceeding $b$, namely $q_{i+1}, \dots, q_{i+b}$. The local circuit depth $D_i^{(AQFT)}$ required to complete all relevant gate operations for this qubit is reduced to $O(b^2)$:
\begin{equation}
D_i^{(AQFT)} = \sum_{k=1}^{\min(b,n-i)} D(CR_k) \le \sum_{k=1}^b O(k) = O(b^2).
\end{equation}
By linearly superimposing the local depths of $n$ qubits, the total circuit depth $D_{AQFT}^{serial}$ of AQFT is obtained:
\begin{equation}
D_{AQFT}^{serial} = \sum_{i=1}^n D_i^{(AQFT)} \le \sum_{i=1}^n O(b^2) = n \cdot O(b^2).
\end{equation}
That is, $O(nb^2)$. In practical simulations, to balance precision and resources, the truncation parameter $b$ is selected as a fixed constant, so $b^2$ can be regarded as a constant coefficient, reducing the circuit depth to $O(n)$. If the total error must be confined within an acceptable precision $\epsilon$, the circuit depth can only be reduced to $O(nlogn)$.

Although the aforementioned truncation operation reduces resource consumption, our direct removal of long-range $CR_k$ gates introduces deterministic algorithmic errors. The removed long-range $CR_k$ gates would have imposed a minor yet cumulative phase shift on the target qubit. Without additional intervention, this phase shift would disrupt the coherent interference of quantum states, potentially causing non-physical deviations in the high-frequency momentum components of the fluid after it is mapped to momentum space.

To compensate for this loss of precision, we introduced a single-qubit gate phase compensation operation. The core idea is to use single-qubit gates to approximate the removed two-qubit gates, thereby achieving partial phase compensation. Suppose we remove a $CR_k$ gate with control bit $q_c$ and target bit $q_t$, which adds a phase of $e^{i2\pi/2^k}$ to the $|1\rangle$ state of $q_t$ when $q_c$ is in the $|1\rangle$ state. The compensation strategy involves using a single-qubit $R_z(\theta_{comp})$ gate on the target bit $q_t$ to approximate the average phase effect, where the compensation angle $\theta_{comp}$ is based on the statistical average assumption of the control bit's state.

Let the probability of control bit $q_c$ being in the $|1\rangle$ state be $p_c$. The expected phase shift induced by the removed $CR_k$ gate on the target bit is $\theta_{expect} = p_c \cdot 2\pi/2^k$. In Hamiltonian simulation of fluid dynamics, without prior information on flow field polarization, we can assume the average probability of control qubits being in the $|1\rangle$ state is $p_c \approx 1/2$. The total phase compensation for all removed control qubits ($k > b$) ultimately applied to the target qubit $q_t$ can be superposed and calculated. Through this operation, we approximately corrected algorithmic bias at an extremely low physical cost without adding additional two-qubit gates.

\subsection{Momentum Operator Truncation}
In momentum space, fluid evolution is governed by the operator $U_k(t) = \exp(-i \hat{k}^2 t / 2)$. By leveraging the property of mutual commutativity, the full evolution operator can be decomposed into a product of single-qubit gates $R_z$ and two-qubit gates $R_{zz}$:
\begin{equation}
U_k(t) = e^{-ic_0 t} \left(\prod_{i=0}^{n-1} e^{-ic_i t Z_i}\right) \left(\prod_{i<j}^{n-1} e^{-ic_{ij} t Z_i Z_j}\right).
\end{equation}
Ignoring the first global phase term $\exp(-ic_0t)$, the two-qubit entanglement part relies on $R_{zz}(\theta_{ij})$ gates with corresponding phase angles $\theta_{ij}=2c_{ij}t$. In unoptimized circuits, there exist $n(n-1)/2$ $R_{zz}$ gates, resulting in a circuit depth of $O(n^2)$ for this section. The coefficient $c_{ij}$ increases exponentially with the growth of the indices $i$ and $j$ of the qubits, resulting in an extremely large theoretical phase angle $\theta_{ij}$ for high-frequency qubit pairs.

If $\theta_{ij}$ is truncated directly based on absolute magnitude, high-frequency terms cannot be eliminated. Due to the $2\pi$ periodicity of quantum gate phase evolution, it is necessary to map all phases to the physically meaningful interval $[-\pi, \pi]$. After obtaining the true phase $\tilde{\theta}_{ij}$, an artificial truncation threshold $\epsilon_{th}$ is introduced. The truncation criterion for the $R_{zz}$ gate is defined as:
\begin{equation}
R_{ZZ}(\tilde{\theta}_{ij}) \approx
\begin{cases}
R_{ZZ}(\tilde{\theta}_{ij}), & \text{if } |\tilde{\theta}_{ij}| \ge \epsilon_{th} \\
I, & \text{if } |\tilde{\theta}_{ij}| < \epsilon_{th}
\end{cases}.
\end{equation}
When the absolute value of the phase is greater than or equal to the threshold, it indicates that the entanglement gate plays a significant role in the evolution of fluid momentum and should be retained in the circuit. Conversely, when the absolute value of the physical phase is below the threshold, it is approximated as $I$, and we directly remove the $R_{zz}$ gate from the quantum circuit.

Extracting the entangled two-qubit interaction term from the wave number operator formula reveals that for any quantum bit pair $(i, j)$, the physical phase corresponding to the $Z_iZ_j$ interaction is:
\begin{equation}
\theta_{ij}=-\frac{t}{2}\cdot\frac{2\cdot2^{i+j}}{4}=-t\cdot2^{i+j-2}
\end{equation}

After introducing truncation, assuming the evolution time step is a rational multiple of $\pi$, let $t=\pi/2p$, where $p$ is a constant. At this point, the theoretical phase becomes:
\begin{equation}
\theta_{ij} = -\pi \cdot 2^{i+j-p-2}.
\end{equation}
Entangling gates that satisfy either of two conditions can be removed: Condition 1 requires the phase modulo $2\pi$ to be zero, while Condition 2 requires the phase to be below a threshold. We define the specifics of retaining entangling gates based on these two criteria.

First is the condition where the phase modulo $2\pi$ is zero, since the quantum phase exhibits $2\pi$ periodicity. When $\theta_{ij}$ is an integer multiple of $2\pi$, the $R_{zz}$ gate becomes physically equivalent to the identity operator $I$. Consequently, removing this $R_{zz}$ gate introduces no error, provided the following condition is met:
\begin{equation}
2^{i+j-p-2} \ge 2 \implies i+j-p-2 \ge 1 \implies i+j \ge p+3.
\end{equation}
The qubit indices satisfy $i+j \ge p+3$.

Secondly, if the phase is below the threshold, we introduce a minimal tolerance threshold $\epsilon$. If $|\theta_{ij}|$ is less than $\epsilon$, the $R_{zz}$ gate should be removed from the circuit, provided the following condition is met:
\begin{equation}
\pi \cdot 2^{i+j-p-2} < \epsilon \implies 2^{i+j-p-2} < \frac{\epsilon}{\pi} \implies i+j < \log_2\left(\frac{\epsilon}{\pi}\right) + p + 2.
\end{equation}
That is, the qubit indices satisfy $i + j < \log_2(\epsilon/\pi) + p + 2$.

Combining the two directions mentioned above, if the $R_{zz}$ gate can be preserved in the circuit, its index must meet the following requirements:
\begin{equation}
\log_2\left(\frac{\epsilon}{\pi}\right) + p + 2 \le i+j < p+3.
\end{equation}
Let the width of this interval be $\Delta$, so the width is:
\begin{equation}
\Delta=(p+3)-\left[\log_2\left(\frac{\epsilon}{\pi}\right)+p+2\right]=1-\log_2\left(\frac{\epsilon}{\pi}\right)
\end{equation}
The interval width $\Delta$ depends solely on the time parameter $p$ and the threshold $\epsilon$, and is independent of the number of qubits $n$. Consequently, the total number of retained two-qubit gates is constrained as follows:
\begin{equation}
N_{gates} \le \frac{1}{2} \sum_{i=0}^{n-1} \Delta = \frac{\Delta}{2} \cdot n.
\end{equation}
Since $\Delta/2$ is a constant, the number of two-qubit gates and the uncompiled circuit depth for the momentum operator evolution part are reduced from $O(n^2)$ to $O(n)$. 

Based on the content of the AQFT part, since the quantum circuit depth of the momentum operator evolution part can be reduced to $O(n)$, and the AQFT part can reduce the circuit depth to $O(nlogn)$ or $O(n)$ using different methods, the overall quantum circuit depth is determined by the circuit depth of the AQFT part.

\subsection{Total Error of Quantum Circuit}
In the approximate optimization scheme proposed in this work, the total theoretical error of the quantum circuit mainly comes from two core truncation operations: the approximate quantum Fourier transform and the truncation of the momentum evolution operator. Operation inevitably introduces a deterministic, purely algorithmic error, and the theoretical upper bound of this error extends as the size of the system increases. 

The first source of total algorithm error is the AQFT module. In the process of state preparation and momentum space conversion, the circuit removes the long-range controlled phase gate $CR_K $ with physical distance $k>b $ by setting the truncation threshold $b$. Although we introduce single-qubit gates for phase compensation in the following steps to approximately compensate for the average phase effect, the basic phase deviation caused by the abandonment of long-range control gates will gradually accumulate. From the theoretical scale, the algorithm error introduced by the AQFT module shows a linear growth trend with the number of qubits, that is, $O(n)$. 

The second part of the significant error comes from the truncation of the momentum space evolution operator. When dealing with pairwise qubit entanglement, we set a tolerance threshold $\ epsilon $, which directly removes $R_{ZZ } $ entanglement gates whose physical phase absolute value is below this threshold. With the expansion of the system size, the number of truncation of the high-frequency bit-pair coupling terms in the momentum space of the fluid increases dramatically, and the theoretical errors introduced by them show an accelerated upward trend. The growth scale of this part of the momentum truncation error is $O (n^2) $ at the quadratic level. 

To sum up, at a fixed truncation threshold, the total algorithmic theoretical error of the quantum circuit is composed of the linear error component of AQFT and the quadratic error component of momentum truncation. However, in practical engineering applications, the evaluation of the fidelity of the overall simulation must take into account the theoretical algorithm error and the physical cumulative error of the hardware. Two-qubit gates of the order of $O (n ^ 2) $in standard circuits will induce catastrophic gate error accumulation on NISQ devices if the above truncation is not performed.

\section{Experiment}\label{sec:experiment}

We use 10 qubits to simulate the two-dimensional unsteady divergent flow on the Songshan Supercomputer of the National Supercomputing Center in Zhengzhou, using a single node, 16 core CPU, 48G memory, 4 DCU accelerator card. The following shows the evolution results of the divergent flow at three time points: $t=0$, $t=\pi/4$, and $t=\pi/2$, before and after applying the optimization strategy.

\begin{figure}[h]
\centering
\includegraphics[width=0.9\textwidth]{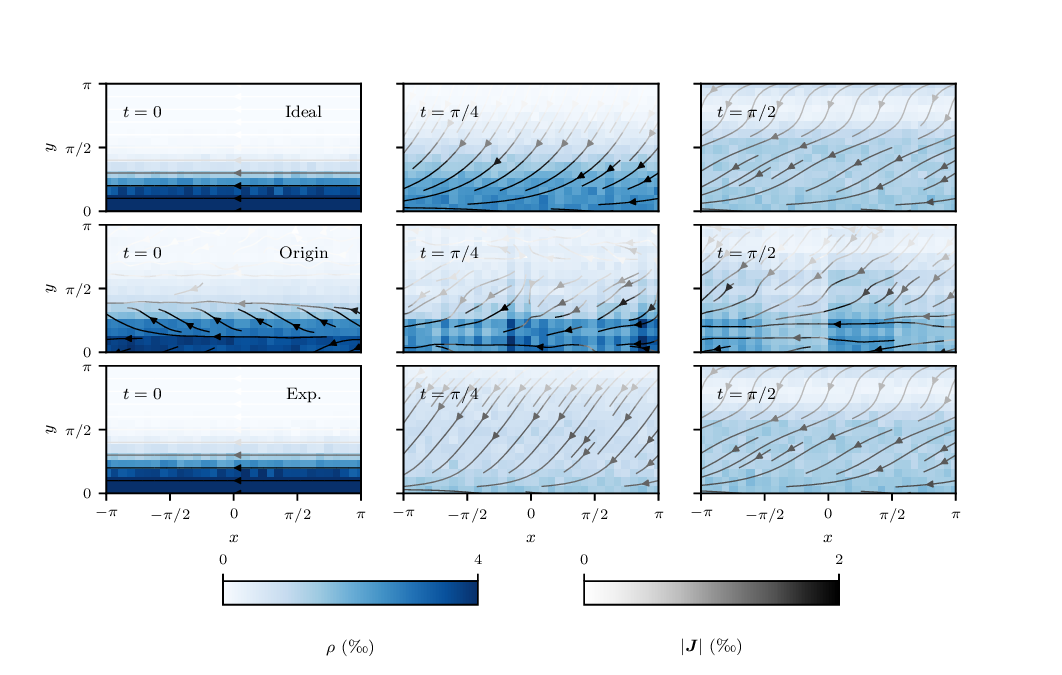}
\caption{Experimental measurement data of density contours and streamlines evolving over time.}\label{Fig1}
\end{figure}

Figure~\ref{Fig1} shows the experimental data of density contours and streamlines evolving over time. The Ideal part is the experimental results of classical simulation, the Origin part is the experimental results of the original quantum simulation without optimization, and the Exp part is the experimental results of the quantum simulation after truncation optimization. We compare the other two parts for truncation optimization. In the process of evolution, the original quantum simulation has serious hardware gate error accumulation caused by deep quantum circuits, and the obtained results have the phenomena of non-physical space oscillation and streamline distortion. The optimized flow field retains the physical characteristics of mass and momentum diffusion to both sides, but due to the truncation of the high-frequency bit coupling term, it will inevitably lead to minor deviations in the fine spatial scale, particularly at the $t=\pi/4$ time step, though the overall pointwise fidelity remains robust with correlation coefficients of $r \ge 0.933$, and the optimized flow field will have some stiff phenomena in the edge transition and diffusion areas, losing the extremely continuous smoothness before optimization.

\begin{figure}[h]
\centering
\includegraphics[width=0.9\textwidth]{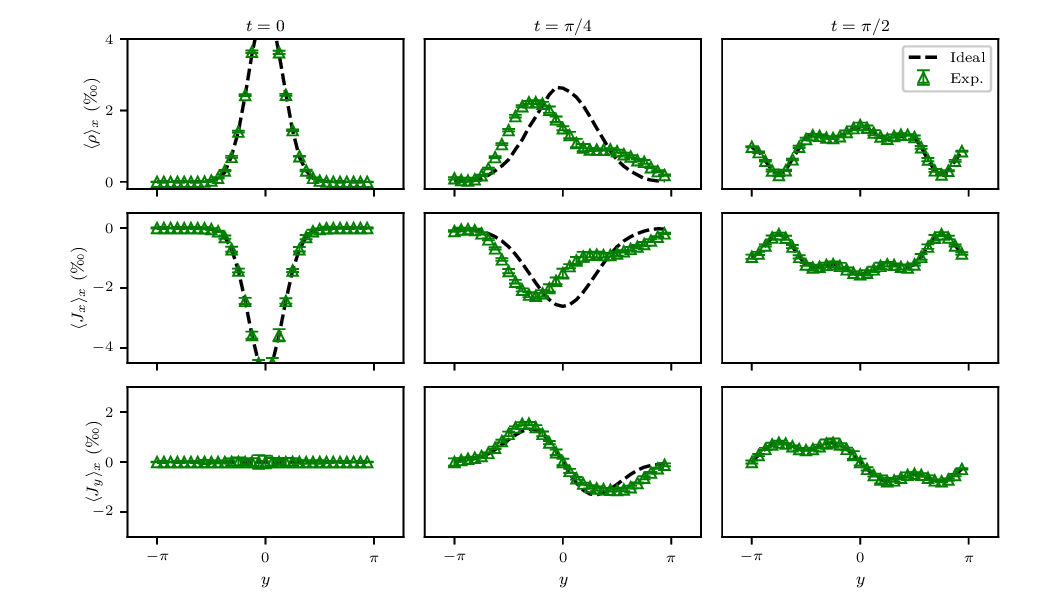}
\caption{Distribution profiles of density $\rho$ and the two components of momentum, $J_x$ and $J_y$.}\label{Fig2}
\end{figure}

Figure~\ref{Fig2} shows the distribution profiles of density $\rho$ and the two components of momentum, $J_x$ and $J_y$. At all times, the optimized profile data closely aligns with the unoptimized baseline data, demonstrating that the optimization strategy reduces line depth without disrupting the macroscopic hydrodynamic evolution of the system. However, in the one-dimensional profile at $t=\pi/4$, a slight deviation from the smooth baseline is observed near the central region at $y=0$, indicating that truncation error manifests more noticeably during evolution at $t=\pi/4$ compared to other times.

\begin{figure}[htbp]
    \centering
    \begin{subfigure}{0.32\textwidth}
        \centering
        \includegraphics[width=\textwidth, height=\textwidth]{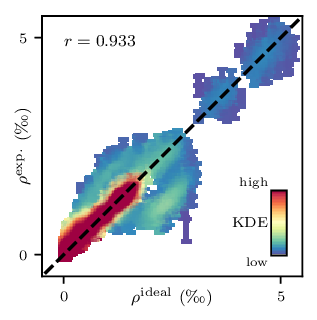}
        \label{fig:sub1}
    \end{subfigure}\hfill
    \begin{subfigure}{0.32\textwidth}
        \centering
        \includegraphics[width=\textwidth, height=\textwidth]{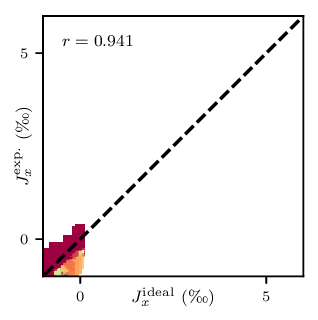}
        \label{fig:sub2}
    \end{subfigure}\hfill
    \begin{subfigure}{0.32\textwidth}
        \centering
        \includegraphics[width=\textwidth, height=\textwidth]{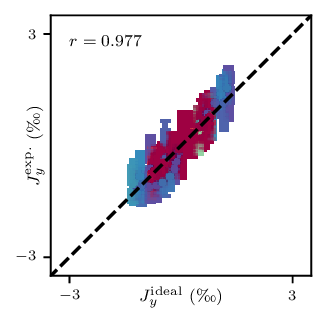}
        \label{fig:sub3}
    \end{subfigure}
    
    \caption{Scatter plots comparing the ideal values versus experimental values of $\rho$, $J_x$, and $J_y$ at different time steps.}
    \label{Fig3}
\end{figure}

Figure~\ref{Fig3} shows scatter plots comparing the ideal values versus experimental values of $\rho$, $J_x$, and $J_y$ at different time steps. The optimized density $\rho^{\text{exp}}$ achieved a correlation coefficient of $r=0.933$ compared to the unoptimized results, the x-direction momentum $J_x^{\text{exp}}$ reached $r=0.941$, and the y-direction momentum $J_y^{\text{exp}}$ achieved $r=0.977$. Since the absolute value of the $J_y$ component is relatively small in this divergent flow model, it is highly susceptible to being overwhelmed or distorted by accumulated hardware noise from two-qubit gates in unoptimized circuits. The correlation coefficient of $0.977$, along with the highly concentrated scatter distribution near the diagonal, strongly demonstrates the effectiveness of this strategy in extracting and preserving weak hydrodynamic features. Truncation directly discards a significant amount of information representing weak quantum entanglement. Although we maintained the macroscopic structure of the flow field, this came at the cost of sacrificing numerical accuracy at individual grid points, leading to slight degradation in pointwise fidelity, which remains bounded with a minimum correlation coefficient of $r = 0.933$.

\begin{figure}[h]
\centering
\includegraphics[width=0.9\textwidth]{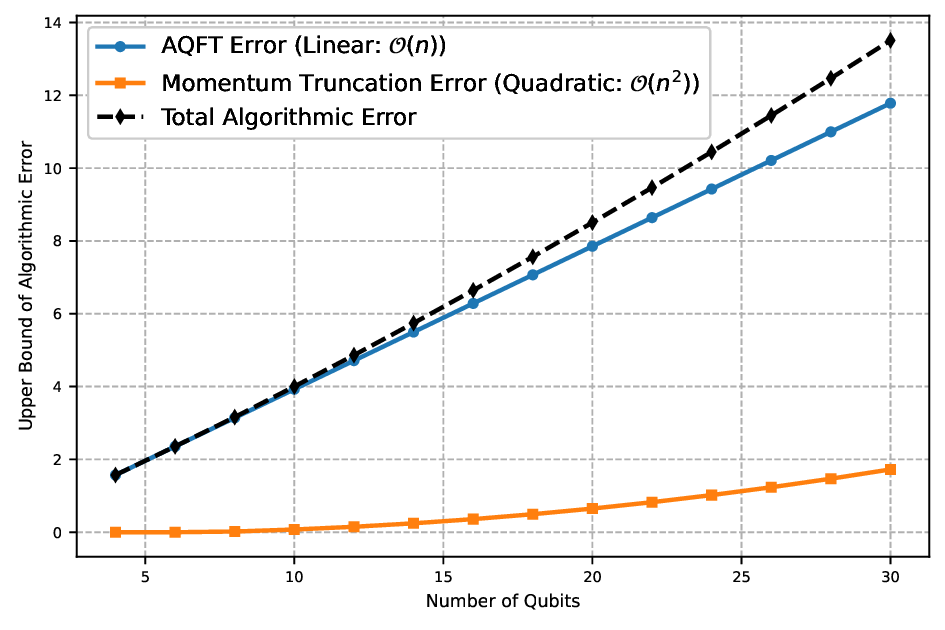}
\caption{Theoretical scaling of algorithmic truncation errors with the number of qubits under fixed thresholds.}\label{Fig4}
\end{figure}

Figure~\ref{Fig4} shows the evolution of the algorithm truncation error as the number of qubits increases at a fixed threshold. The two main sources of total algorithm error are clearly shown in the figure. The blue curve corresponds to the approximate quantum Fourier transform error, which shows a linear growth trend of \textit{O(n)}. This reflects the accumulation of basic phase deviation caused by abandoning the long-range controlled phase gate in the process of transformation between coordinate system and momentum system. The orange curve corresponds to the momentum operator truncation error, which increases by an order of \textit{O(n²)}. With the expansion of the system size, the error introduced by the high-frequency bit coupling term in the truncated momentum space shows an accelerating upward trend. From the point of view of the rigor of mathematical derivation, the pure algorithm error introduced by single-step truncation will proliferate in the depth circuit.

\section{Discussion}\label{sec13}

\begin{figure}[h]
\centering
\includegraphics[width=0.9\textwidth]{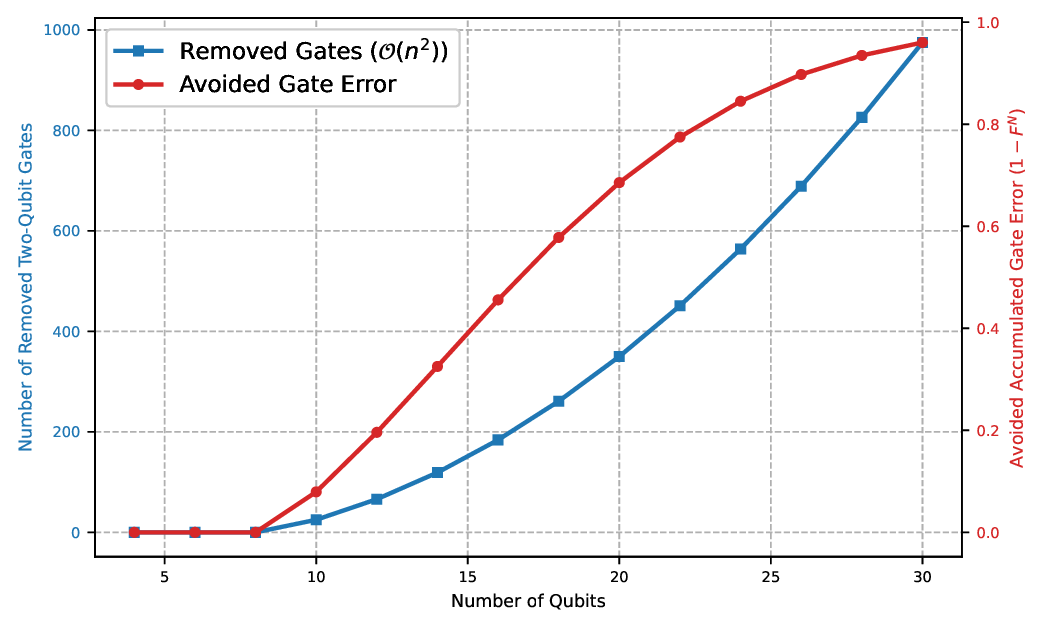}
\caption{Reduction in two-qubit gates and the avoided accumulated hardware error via truncation optimization(Two-Qubit Gate Fidelity = 99.67\%).}
\label{Fig5}
\end{figure}

Figure~\ref{Fig5} reveals that the truncation optimization reduces the number of two-qubit gates as well as the circumvented hardware accumulation error case under the condition that the two-qubit gate fidelity is 99.67\%. The evolution of the two key curves with the number of qubits is shown in the figure. The blue curve corresponds to the number of double-bit gates removed, and the curve clearly indicates that the number of gates removed increases at the level of \textit{O(n²)}. In the standard quantum fluid simulation, the standard quantum Fourier transform module and the momentum operator evolution part require \textit{O(n²)} order of magnitude of two-qubit gates. With the improvement of the simulation grid resolution, that is, the increase of the number of qubits, the optimization strategy accurately strips off those redundant long-range gate operations that explode quadratically with the scale. The deterioration of the depth of the quantum circuit is fundamentally curbed. The red curve corresponds to the evaded cumulative gate error, and the right vertical axis represents the evaded cumulative hardware error under the benchmark of 99.67\% two-qubit gate fidelity for real superconducting physical devices. The data clearly show that the red curve rapidly approaches 100\% error rate when the system is extended to 20 to 30 qubits. This means that if you stick to unoptimized standard circuits, the accumulated errors in the physical hardware will inevitably lead to complete decoherence of the system.

\begin{figure}[h]
\centering
\includegraphics[width=0.9\textwidth]{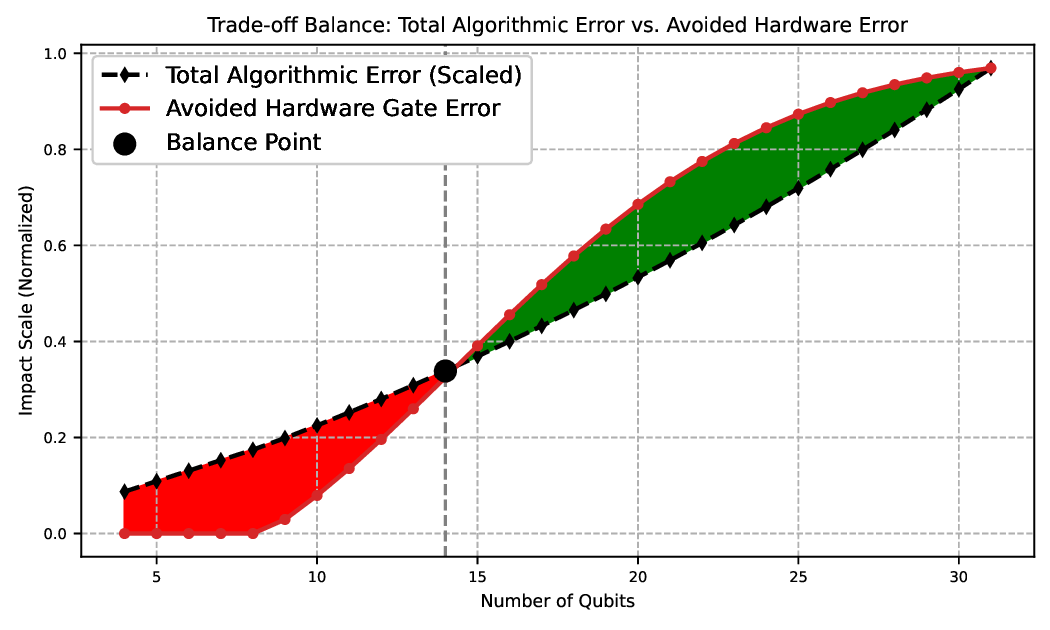}
\caption{Dynamic trade-off analysis between algorithmic truncation error and avoided hardware gate error.}
\label{Fig6}
\end{figure}

Figure~\ref{Fig6} presents a dynamic trade-off analysis between algorithm truncation error and avoided hardware gate error. The red curve corresponds to the avoided hardware gate error. As the number of qubits increases, the cumulative error generated by the hardware will approach 100\% if it is not truncated. The purple curve corresponds to the algorithm (truncation) error. As the scale increases, the theoretical error due to truncation also increases steadily, bounded by a combined growth rate of $O(n)$ from AQFT and $O(n^2)$ from momentum truncation. The intersection of the two curves is the equilibrium point.

Introducing approximation operations in quantum fluid simulations reduces the computational resource pressure by explicitly eliminating the $O(n^2)$ growth of redundant long-range gate operations, but this optimization method essentially trades algorithmic accuracy for physical hardware resources. As an approximate quantum optimization approach, the specific value of the threshold $\epsilon$ significantly impacts the simulation of flow fields. 

If the $\epsilon$ value is set too large for extreme resource optimization, it will erase the high-wavenumber components in the flow field while also diluting the originally sharp density fronts and extreme value regions. Once minor phase gates are indiscriminately discarded, the simulated flow field will become blurred and completely lose its ability to depict fine vortex structures and local large gradients. From a mathematical derivation perspective, the pure algorithmic error introduced by single-step truncation will surge. At this point, the accumulation of algorithmic errors will cause the theoretical fidelity $F$ to rapidly approach zero, and the measured output will degenerate into physically meaningless random white noise.

On the contrary, if $\epsilon$ is conservatively set to retain an excessive number of minimal controlled-phase gates, the momentum evolution circuit remains densely structured, with circuit depth and the number of two-qubit gates still being very high. This approach entirely defeats the purpose of introducing approximate optimization. In ideal noiseless theoretical simulations, this method can perfectly preserve all high-frequency features of the flow field. However, in the case of current NISQ devices, too many gate operations will inevitably bring about hardware cumulative errors that rapidly approach a 100\% error rate when the system scales to 20-30 qubits under a baseline two-qubit gate fidelity of 99.67\%

\section{Conclusion}\label{sec12}

In summary, the key to balancing algorithmic errors in quantum fluid simulations and their computational resource usage lies in selecting the appropriate value of the truncation threshold $\epsilon$. In practical applications, to better reduce the resources used in quantum simulations, $\epsilon$ should not be fixed as a constant but rather treated as a hyperparameter for regulating resource allocation. By controlling the value of the threshold in approximation operations, we can evade near 100\% accumulated hardware errors and achieve scalable fluid simulations with $O(n)$ or $O(nlogn)$ depth within limited quantum computing resources.

The joint optimization strategy for approximate quantum Fourier transform and momentum operator truncation proposed in this paper not only provides a scalable resource compression scheme that explicitly avoids $O(n^2)$ gate accumulations for quantum fluid simulations on noisy intermediate-scale quantum devices, but its core concept can also be widely extended to physics and engineering computations. Within the scope of fluid dynamics, this approach of transforming high-frequency algorithmic errors and hardware noise into utilizable resources can be functionally applied to simulate small-scale turbulent motions, among other aspects. Furthermore, this strategy essentially involves deep tailoring of partial differential equations solved through Hamiltonian-based simulations, making it equally applicable to other computational scenarios heavily reliant on spatial and frequency domain transformations.

Future research can deeply extend this controlled truncation approach to two-dimensional vortex simulations. Meanwhile, general fluid dynamics exhibits nonlinear and non-Hermitian Hamiltonian characteristics. In future work, the computational resource budget liberated by truncation strategies will enable the introduction of high-dimensional mappings within finite coherence times to address complex nonlinear fluid mechanisms. Furthermore, combining variational quantum algorithms for efficient preparation of initial quantum states and further optimizing quantum circuits at the compilation level constitute core directions for improving this technical system. Prior to the full realization of quantum error correction, this methodology of deeply coupling approximate algorithms with underlying hardware characteristics can accelerate the practical advantages of quantum computing in engineering computational fluid dynamics.

\section*{Acknowledge}

This work is supported by the National Key Research and Development Program (2024YFB4504103) and the simulations in this work were implemented on the SongShan supercomputer at National Supercomputing Center in Zhengzhou. This work is also supported by Jiangsu Province Engineering Research Center of IntelliSense Technology and System.

\bibliographystyle{unsrt}
\bibliography{bibliography}

\end{document}